\documentclass[aps,pra,notitlepage,superscriptaddress,twocolumn,floatfix]{revtex4-2}
\usepackage{amsmath,amssymb,amsfonts,bm,comment}
\usepackage{graphics,float,xcolor}
\usepackage[caption=false]{subfig}
\usepackage[colorlinks=true,
linkcolor=blue,
filecolor=magenta,      
urlcolor=blue,
citecolor=blue]{hyperref}
\usepackage{physics,xparse,mathtools,isotope}
\usepackage[note-name = ,use-sort-key = false]{notes2bib}

\usepackage[table-parse-only]{siunitx}
\newcommand{\im}{\mathrm{i}}

\newcommand{\figref}[1]{\unskip\;\textup{\ref{#1}}}
\newcommand{\dash}{\mbox{--}}

\makeatletter
\let\origref\ref
\renewcommand{\eqref}[1]{\unskip\;\textup{\tagform@{\origref{#1}}}}
\makeatother
\graphicspath{{figures/}}

\begin{document}
\title{Vortex formation and quantum turbulence with rotating paddle potentials in a two-dimensional binary Bose-Einstein condensate}
\author{Subrata Das}
\email{subratappt@iitkgp.ac.in}
\affiliation{Department of Physics, Indian Institute of Technology Kharagpur, India}
\author{Koushik Mukherjee}
\affiliation{Department of Physics, Indian Institute of Technology Kharagpur, India}
\author{Sonjoy Majumder}
\email{sonjoym@phy.iitkgp.ac.in}
\affiliation{Department of Physics, Indian Institute of Technology Kharagpur, India}
\date{\today}

\begin{abstract}
    We conduct a theoretical study of the creation and dynamics of vortices in a two-dimensional binary Bose-Einstein condensate with a mass imbalance between the species. To initiate the dynamics, we use one or two rotating paddle potentials in one species, while the other species is influenced only via the interspecies interaction. In both species, the number as well as the dominant sign of the vortices are determined by the rotation frequency of the paddle potential. Clusters of positive and negative vortices form at a low rotation frequency comparable to that of the trap when using the single paddle potential. In contrast, vortices of the same sign tend to dominate as the rotation frequency of the paddle increases, and the angular momentum reaches a maximum value at a paddle frequency, where the paddle velocity becomes equal to the sound velocity of the condensate. When the rotation frequency is sufficiently high, the rapid annihilation of vortex-antivortex pairs significantly reduces the number of vortices and antivortices in the system. For two paddle potentials rotating in the same direction, the vortex dynamics phenomenon is similar to that of a single paddle. However, when the paddle potentials are rotated in the opposite direction, both positive and negative signed vortices occur at all rotational frequencies. At the low rotation frequencies, the cluster of like-signed vortices produces the $k^{-5/3}$ and $k^{-3}$ power laws in the incompressible kinetic energy spectrum at low and high wavenumbers, respectively, a hallmark property of the quantum turbulent flows.
\end{abstract}

\maketitle
\section{Introduction}\label{sec:intro}
Vortices \cite{simula_2019_quantised} render Bose-Einstein condensate (BEC) an excellent platform for examining various scaling aspects of quantum turbulence \cite{allen_2014_quantum,madeira_2019_quantum,madeira_2020_quantum} which are quantum counterparts of classical turbulence \cite{holmes_2012_turbulence,onsager_1949_statistical,eyink_2006_onsager}. The renowned Kolmogorov's `$5/3$' law is one of the most well-known of these scaling laws among them~\cite{nore_1997_kolmogorov,kobayashi_2005_kolmogorov}. Several strategies are available to the current state BEC experiments \cite{anderson_2010_resource} to generate non-linear defects such as vortices and solitons \cite{matthews_1999_vortices,anderson_2000_vortex,chai_2020_magnetic,lannig_2020_collisions,katsimiga_2020_observation,navarro_2013_dynamics}.
These include laser stirring \cite{inouye_2001_observation,raman_2001_vortex,neely_2010_observation}, rotating the confining potential \cite{hodby_2001_vortex,williams_2010_observation,hodby_2003_experimental}, interaction with the optical vortex \cite{mondal_2014_angular,bhowmik_2016_interaction,das_2020_transfer,mukherjee_2021_dynamics}, quenching through the phase transition \cite{zurek_1985_cosmological}, and counter-flow dynamics \cite{carretero-gonzalez_2008_dynamics,xiong_2013_distortion,yang_2013_dynamical,mukherjee_2020_quench}, just to name a few \cite{tsubota_2002_vortex,leanhardt_2002_imprinting,kumakura_2006_topological}.
Theoretically, numerous intriguing aspects of three-dimensional \cite{nore_1997_kolmogorov,wells_2015_vortex,navon_2019_synthetic,garcia-orozco_2021_universal,serafini_2017_vortex} and two-dimensional (2D) quantum turbulence (QT) \cite{henn_2009_emergence,horng_2009_twodimensional,white_2010_nonclassical,numasato_2010_direct,bradley_2012_energy,reeves_2013_inverse,villasenor_2014_quantum,billam_2014_onsagerkraichnan,stagg_2015_generation,mithun_2021_decay,estrada_2022_turbulence,dasilva_2022_vortex} have been examined. Moreover, very recently developed machine learning techniques can be utilized to detect and classify quantum vortices \cite{metz_2021_deeplearningbased,sharma_2022_machinelearning}.
The incredible tunability of atom-atom interaction via Feshbach resonance \cite{chin_2010_feshbach,kohler_2006_production}, as well as the outstanding maneuverability of dimension \cite{gorlitz_2001_realization}, have also resulted in the significant development of the QT experiment in BEC. In that regard, Ref. \cite{henn_2009_emergence} shows a turbulent tangle of vortices formed by oscillating perturbation. Spontaneous clustering of the same circulation vortices has also been demonstrated experimentally \cite{gauthier_2019_giant,johnstone_2019_evolution}. It is worth noting that clustering of vortices \cite{yu_2016_theory,reeves_2014_signatures} implies the transfer of energy from small to large length scales, illustrating the so-called inverse energy cascade \cite{navon_2019_synthetic,navon_2016_emergence}, a well-known phenomenon that occurs in classical 2D turbulence \cite{kraichnan_1967_inertial,kraichnan_1975_statistical}.
The experiment in \cite{johnstone_2019_evolution}, for instance, employs a paddle that swifts through the bulk of the BEC, causing randomly distributed vortices that fast assemble into Onsager point vortex clusters, a notion that has also been theoretically studied by White \emph{et al.} \cite{white_2012_creation}.\par
Given that optical paddle potential is a dependable way to create 2D QT, we attempted to conduct a detailed theoretical examination of the production of vortex complexes, the behaviour of angular momentum and the onset of quantum turbulence in a two-component system by utilizing rotating paddle potentials.
Furthermore, we use a more complicated system with 2D binary BECs  \cite{papp_2008_tunable,wang_2015_double,mccarron_2011_dualspecies}, where only one species is exposed to the rotating paddle. We specifically identify the frequency regimes of the rotating paddle where the maximum angular momentum can be imparted to the condensates, as well as systemically investigate the distinct behavior emerging from single and double paddle potential.
The system of 2D binary BECs, which exhibits a variety of instability phenomena \cite{maity_2020_parametrically,sasaki_2009_rayleightaylor,gautam_2010_rayleightaylor,suzuki_2010_crossover,ruban_2022_instabilities} and non-linear structures \cite{mueller_2002_twocomponent,schweikhard_2004_vortexlattice,kuopanportti_2012_exotic,kasamatsu_2009_vortex} is intriguing on its own right.
Using the so-called tune-out technique \cite{leblanc_2007_speciesspecific}, the previously mentioned species selective interaction resulting in the formation of optical paddle potential can be experimentally performed.
In this tune-out method, when one species interacts with the laser light, the other remains unaffected. Furthermore, we investigate a wide range of rotating frequencies of the paddle potential, allowing us to pinpoint the domain in which clustering of the same circulation vortices arises, exhibiting the well-known scaling rule of 2D QT \emph{i.e.} Kolmogorov's $-5/3$ scaling law \cite{reeves_2012_classical, bradley_2012_energy, mithun_2021_decay}. Although different stirring configurations are available in the literature \cite{sasaki_2010_benard,parker_2005_emergence,white_2012_creation,gauthier_2019_giant,muller_2022_critical}, the main objective of employing a rotating
paddle potential in this manuscript is to impart a finite net angular momentum to one of the
binary species within a specific frequency regime and transfer this momentum to the other
species.
We also look at a region dominated solely by identical signed multiple vortices. Furthermore, when the paddle rotates more vigorously, the vortical content of the system drops due to the generation of a high amount of sound waves \cite{leadbeater_2001_sound,parker_2005_emergence,horng_2009_twodimensional,simula_2014_emergence}.
When there is finite interspecies contact interaction, vortex formation can occur even in the second component of the condensate. Most importantly, the vortex in one component is connected by a complementary structure, referred to as a vortex-bright soliton \cite{law_2010_stable,mithun_2021_decay},  in the other.
Besides, we demonstrate the effect of double paddle potentials, in which paddles can rotate either in the same or opposite directions.\par
This article is arranged as follows.
Sec. \ref{sec:gp} describes our setup and delves over the Gross-Pitaevskii (GP) equations.
In Sec. \ref{sec:vor_dyna}, we investigate the non-equilibrium dynamics of a binary system consisting of a mass-imbalanced system using both single (Sec. \ref{sec:single_p}) and double paddle potential (Sec. \ref{sec:double_p}).
Section \ref{sec:energy_spec} examines the incompressible and compressible kinetic energy spectra.
Finally, we summarise our findings and discuss potential future extensions in Sec. \ref{conclusion}.
Appendix \ref{sec:equal_mass} briefly describes the creation of vortices and their dynamics in a binary BEC with equal mass. In  Appendix \ref{sec:negative_paddle}, we demonstrate vortex creation using the negative paddle potential.

\section{Gross-Pitaevskii equation}\label{sec:gp}
We consider binary BECs, referred to as species A and B, that are confined in 2D harmonic trapping potentials \cite{kwon_2021_spontaneous}. The species consists of $N_i$ number of atoms of mass $m_i$ ($i=\rm A,B$). The form of the trapping potentials read $V_{\rm trap} = \frac{1}{2}m(\omega^2_x x^2 + \omega^2_y y^2 + \omega^2_z z^2)$, where $\omega_x, \omega_y$ and $\omega_z$ are trapping frequencies along $x,y$ and $z$ directions, respectively. To implement a quasi-2D BEC in the $x$-$y$ plane, we consider the following criterion for the trap frequencies, namely, $\omega_x= \omega_y=\omega \ll \omega_z$. We apply single or double stirring potential $V_P$ generated by a far-off-resonance blue-detuned laser beam shaped into an elliptic paddle in species A to induce vortices in the condensate~\cite{gauthier_2019_giant}. The potential $V_{P_{\alpha}}$, with $\alpha \in \{1,2\} $ can be expressed as \cite{white_2012_creation}
\begin{align}\label{eq:paddle_pot}
    V_{P_{\alpha}}(x,y,t) = & V_0 \exp\Big[  -\frac{\eta^2 \qty(\tilde{x}_{\alpha}\cos(\omega_p t)-\tilde{y}_{\alpha}\sin(\omega_p t))^2}{d^2} \nonumber \\  &-\frac{(\tilde{y}_{\alpha}\cos(\omega_p t)+\tilde{x}_{\alpha}\sin(\omega_p t))^2}{d^2} \Big],
\end{align}
where $\tilde{x}_{\alpha}=x - x_{p, \alpha}$ and $\tilde{y}_{\alpha}=y-y_{p, \alpha}$, considering the center of the paddle potential at $(x_{p, \alpha},y_{p, \alpha})$ for the $\alpha$ paddle. Here $V_0$ is the peak strength of the potential, $\omega_p$ is the rotation frequency of the paddle, and $\eta$ and $d$ determine the paddle elongation and width, respectively.

In the quasi-2D regime, the motions of atoms along $z$-direction become insensitive and the wavefunctions $\Psi_{\rm A(B)}$ can be expressed as  $\psi_{\rm A(B)}(x,y)\zeta(z)$, where $\zeta(z)=(\lambda/\pi)^\frac{1}{4} \exp(-\lambda z^2/2)$ is the ground state along $z$ direction and $\lambda=\omega_z/\omega$ is the trap aspect ratio. After integrating out the $z$ variable, the 2D dimensionless time-dependent GP equation that governs the dynamics of a BEC is given by \cite{pethick_2008_bose,pitaevskii_2003_boseeinstein}
\begin{multline}\label{eq:gp}
    \im \pdv{\psi_i}{t}=\Bigg[-\frac{1}{2}\frac{m_{\rm B}}{m_i}\qty(\pdv[2]{}{x}+\pdv[2]{}{y})  +  \frac{1}{2}\frac{m_i}{m_{\rm B}}\qty(x^2 + y^2)  \\ +  \sum_{j= \rm{A,B}} g_{ij}N_j\abs{\psi_j}^2 + \delta_{{\rm A}i}(V_{P_{1}}+ V_{P_{2}})   \Bigg] \psi_i,
\end{multline}
where $i= {\rm A, B}$.
Here, the effective 2D non-linear interaction coefficient is determined by the term $g_{ij}=\sqrt{\lambda/(2\pi)} 2\pi a_{ij}m_{\rm B}/m_{ij}$ with $a_{ij}$ being the scattering length,  $l=\sqrt{\hbar/(m_{\rm B} \omega)}$ is the oscillator length,  $m_{ij}=m_im_j/(m_i+m_j)$ denotes the reduced mass.
The dimensionless Eq. \eqref{eq:gp} is written in terms of length scale $l$, time scale $1/\omega$ and energy scale $\hbar\omega$. The $i$-th species wavefunction is normalized to $\int\abs{\psi_i}^2\dd^2{r}=1$.

In this paper, we explore the turbulent phenomena that arise from the potentials formed by the rotating single paddle, $V_{P_{1}}$ and the double paddles, $ V_{P_1} + V_{P_2}$. The paddle potentials are maintained in the condensate for the time $0 \le t \le \tau$. Afterward, the paddle is linearly ramped off to zero over a time $t=\Delta\tau$,  during which the relation,
\[V_{P_{1(2)}} \rightarrow V_{P_{1(2)}}\qty(1-\frac{t-\tau}{\Delta\tau}), \] holds in the Eq.\eqref{eq:paddle_pot}. Here we consider a binary BEC of $^{133}$Cs (species A) and $^{87}${Rb} (species B) elements having different masses \cite{mccarron_2011_dualspecies}. The number of atoms in both species A and B are equal, and we take  $N_{\rm A}=N_{\rm B}=60000$. The harmonic trap potential is designed to have a frequency of  $\omega=2\pi\times30.832$ rad/s and the aspect ratio $\lambda=100$. The intra-species scattering lengths are $ a_{\rm AA}=280a_0 $ and $ a_{\rm BB}=100.4a_0 $, where $ a_0 $ is the Bohr radius \cite{mccarron_2011_dualspecies}. The interspecies scattering length $ a_{\rm AB} $ is chosen to reside in the miscible regime, as the following relation of the miscibility $ \emph{i.e.}~a^2_{\rm AB} \le a_{\rm AA} a_{\rm BB}$ \cite{ao_1998_binary}, is hold obeyed by the scattering lengths. We numerically solve the GP equation using the Split-step Crank-Nicolson method \cite{muruganandam_2009_fortran}. The ground state of the system is generated by propagating the wavefunctions of the BEC in imaginary time. In order to inspect the dynamical evolution of the condensate, we utilize the ground state generated in imaginary time as the initial state and solve the Eq. \eqref{eq:gp} in real-time. Moreover, the system's initial state is prepared by placing a paddle-shaped stationary obstacle, as expressed in Eq. \eqref{eq:paddle_pot}, in the component $A$. Our simulation runs on the spatial extent of $ -20.48l $ to $ 20.46l $ along both  $x$ and $y$ direction with 2D $2048 \cross 2048$ grid points.

\section{Creation of vortices using paddle potential}\label{sec:vor_dyna}
As discussed in Ref.~\cite{white_2012_creation}, using an optical paddle potential vortex in BEC can be generated in a variety of ways which include $(\rm i)$ rotating the paddle about a fixed center, $(\rm ii)$ moving the paddle about a fixed center, and $(\rm iii)$ both rotating and moving paddle simultaneously in the BEC. Though we have considered only the rotation of paddle potential to generate vortices in this work, we have employed both a single paddle and a double paddle potential to generate a vortex. In particular, while the single paddle potential rotates in species A with the paddle center being located at $(x_p,y_p)=(0,0)$, the double
paddle potentials can rotate either in the same or opposite directions about their center at  $(x_p,y_p)=(\pm r_{\rm A}/4,0)$, respectively, where $r_{\rm A}=6.1l$  is the root-mean-squared radius of species A $ (\text{for } a_{\rm AB}=0)$. The parameters for single paddle are $\eta=0.05 $ and $d=0.1l$; and for double paddle  $\eta=0.1 $ and  $d=0.1l $, are identical for both. These values determine the elliptical shape of the paddle according to Eq. \eqref{eq:paddle_pot}. Moreover, we choose the peak strength of the paddles to be  $ V_0=10\mu_{\rm A} $, where $ \mu_{\rm A} $ is the chemical potential of species A. As previously stated, after establishing the initial state with a stationary paddle, at $ t=0 $, the paddle is rotated at a frequency of $\omega_p$ with full amplitude until the time $ \tau=40\omega^{-1}=206{\rm ms}$ and then ramped off to zero within $\Delta\tau=10\omega^{-1}$. With these parameters $ \omega_p, \eta, d$, and  $V_0$,  the paddle potentials may be externally regulated, allowing for control of the creation of vortex or antivortex in BEC. In BEC, the presence of a vortex or an antivortex yields a finite amount of angular momentum which can be expressed, for $i$-th species, as
\begin{align}\label{eq:ang_mom}
    L_z^i = -\im \int  \psi_i^*\qty(x\pdv{y}-y\pdv{x})\psi_i\dd x \dd y.
\end{align}
To study the dynamical formation of the vortices, we measure
the density-weighted vorticity of condensates as~\cite{mukherjee_2020_quench, ghoshdastidar_2022_pattern}
\begin{align} \label{eq:vorticity}
    \vb{\Omega}_i = \curl{ \vb{J}_i} ,
\end{align}
for a better spatially resolved measurement with $\vb{J}_i = \frac{\im\hbar}{2m}(\psi_i\grad\psi_i^* - \psi_i^*\grad\psi_i)$ being the probability current density. We remark that by using the Madelung transformation \cite{madelung_1927_quantentheorie}, $\psi_i = \sqrt{n_i} e^{\im \phi_i}$, Eq. \eqref{eq:vorticity} can be cast into the form $\vb{\Omega}_i = \nabla \times n_i\vb{u}_i$. Notably, the multiplication of the condensate velocity $\vb{u}_i$ with the density $n_i$ ensures that we compute the vorticity of $i$-th species only in the region where the condensate is located. \newline

\subsection{Single paddle}\label{sec:single_p}
This section examines the implications of a single paddle potential rotating with frequency $ \omega_p $ about the center of the species A. Although rotation orientation can be clockwise (CW) or counter-clockwise (CCW), we focus on a paddle rotating in the CW direction. We note that the{\unskip\parfillskip 0pt \par}
\onecolumngrid
\vspace*{-2mm}
\begin{figure}[H]
    \centering
    \includegraphics[width=\textwidth]{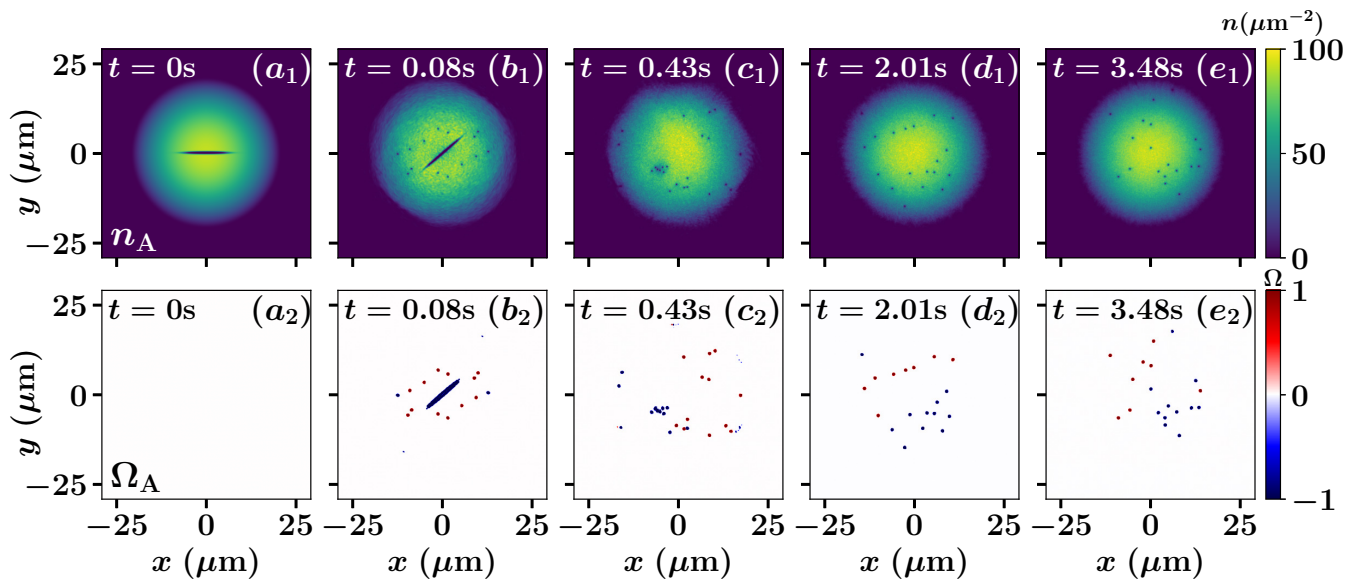}
    \caption{Snapshot of ($a_1$)-($e_1$) density $(n_{\rm A})$ and ($a_2$)-($e_2$) vorticity $(\Omega_{\rm A})$ profiles of the species A at different instants of time (see legends). The binary BECs are made of $^{133}$Cs-$^{87}$Rb atoms. An elliptical paddle potential characterized by the parameters $\eta = 0.05$ and $d=0.1l$ is rotated with the angular frequency $\omega_{p} = \omega$ within the species A ($^{133}$Cs) in order to trigger the dynamics. The colorbars of \textit{top} and \textit{bottom} rows represent the number density ($n$) in  $\mu\rm{m}^{-2}$ and the vorticity ($\Omega$), respectively. The binary BECs are initialized in a two dimensional harmonic potential with frequency $\omega /(2 \pi) = 30.832$ Hz, $\lambda=100$ and having following intra- and interspecies scattering lengths $a_{\rm AA}=280a_{0}$, $a_{\rm BB} = 100.4a_{0}$, and $a_{\rm AB} = 0$. The number of atoms for both the species are $N_{\rm A}=N_{\rm B}=60000$. }
    \label{fig:csrb_den_vor_a0}
\end{figure}
\twocolumngrid
\noindent results obtained for the CCW will be essentially identical to those obtained for the CW.
At first, we demonstrate the behavior of the BEC without interspecies interaction by setting $a_{\rm AB}=0$. Due to the absence of interspecies interactions, the paddle potential does not influence species B, and therefore the latter remains unaltered during the dynamics. When the paddle rotates in species A, vortices and antivortices form around it. The number of vortices and antivortices, in particular, is strongly dependent on the rotation frequency. Figures \figref{fig:csrb_den_vor_a0}($a_1$)-($a_2$) and ($b_1$)-($b_2$) shows time evolution of density and vorticity of species A at the paddle frequency $ \omega_p=\omega $, the trap frequency.

The initial state of species A, with the paddle potential being elongated along the $x$-axis, is shown in the Fig. \figref{fig:csrb_den_vor_a0}($a_1$). At $t = 0.08 s$, after the rotation of the paddle has been established [Fig.\figref{fig:csrb_den_vor_a0}$(b_2)$], both vortices (red color) and antivortices (blue color) are generated  in species A. In fact, a close inspection of the Fig. \figref{fig:csrb_den_vor_a0}$(b_2)$ reveals that the vortex-antivortex structures are located symmetrically  with respect to the paddle. Additionally, the number of vortices exceeds that of the antivortices [Fig. \figref{fig:csrb_den_vor_a0}$(b_2)$].

\begin{figure}[h]
    \includegraphics[width=0.5\textwidth]{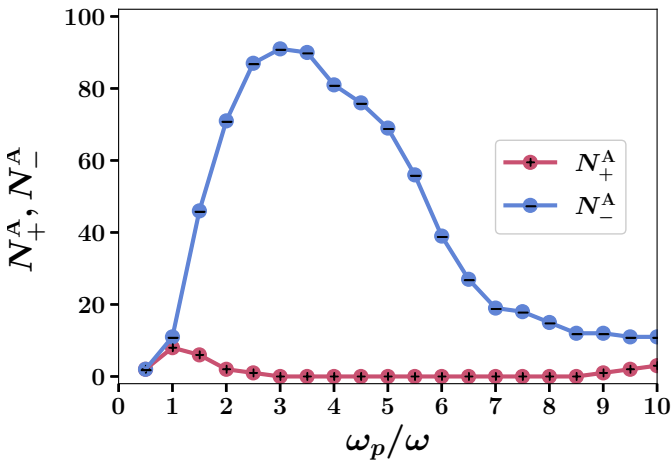}
    \caption{Variation of the number of vortices $N^{\rm A}_+$ and antivortices $N^{\rm A}_-$ of species A at steady state as a function of the paddle frequency $\omega_p$. The other parameters are the same as the ones in Fig. \figref{fig:csrb_den_vor_a0}.}
    \label{fig:vortex_number_a0}
\end{figure}

\begin{figure}[h]
    \includegraphics[width=0.5\textwidth]{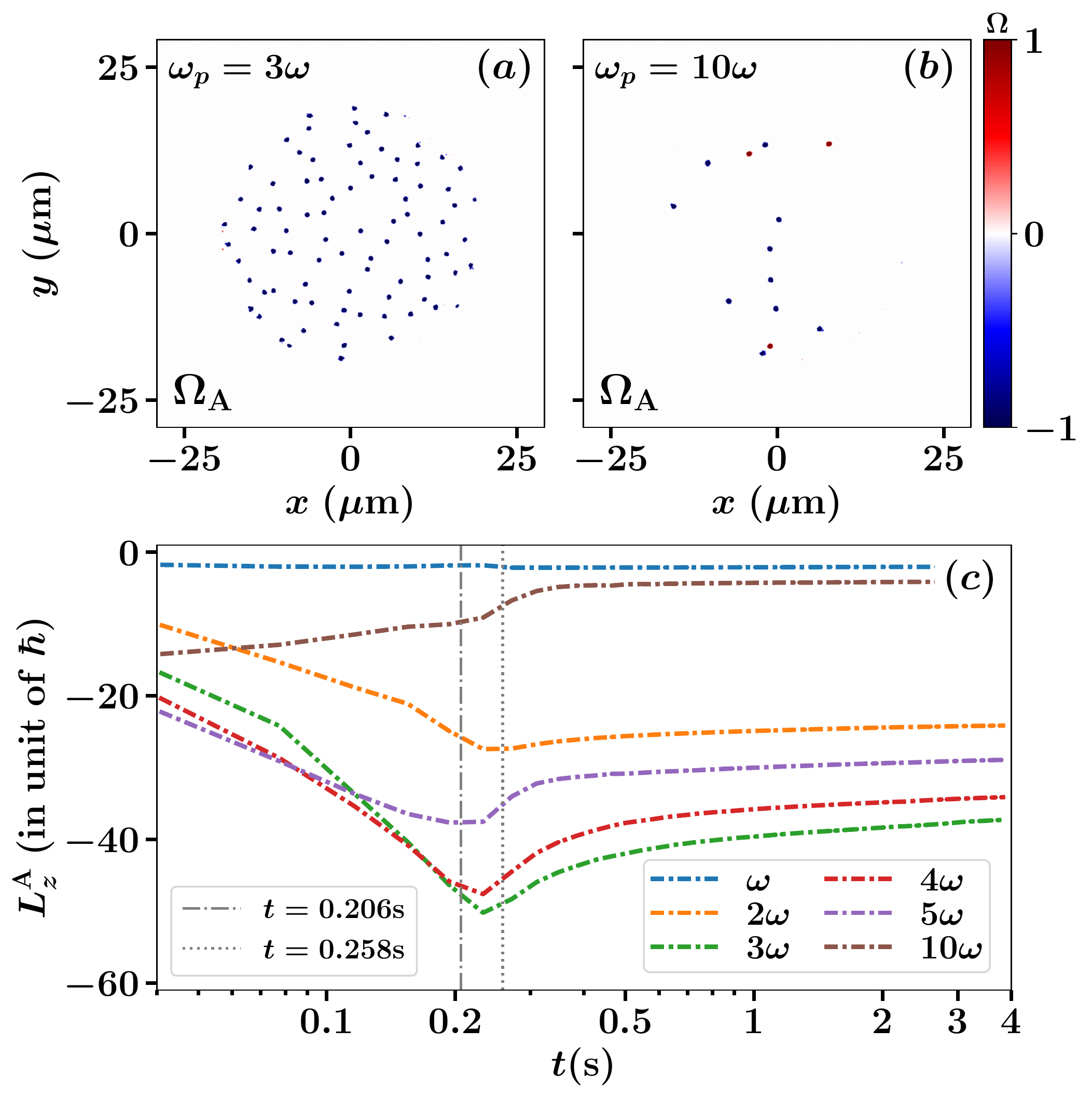}
    \caption{Snapshots of the vorticity profiles of species A $(\Omega_{\rm A})$ taken at $t=3.5\rm s$ for two different paddle frequencies ($a$) $\omega_p=3\omega$ and ($b$) $\omega_{p} = 10\omega$ of rotation of the paddle potential. Shown also $(c)$ the  time-evolution ($\log$-scale) of angular momentum $(L_z^{\rm A})$ for different values of $\omega_p$ (see the legends). The vertical lines in (c) represent the times when amplitude of the paddle started to ramp off and become zero, respectively. The colorbar of \textit{top} row represents the vorticity $(\Omega)$. The other parameters are the same as the ones in Fig. \figref{fig:csrb_den_vor_a0}.}
    \label{fig:csrb_vor_am_a0}
\end{figure}

The generation of vortices and antivortices continues until $ 0.258 \rm s $, at which point the paddle potential vanishes. It is worth noting that the numbers of vortices and antivortices are nearly equal around this time. Following that, a considerable number of the vortex-antivortex pairs decay due to self-annihilation or drifting out of the condensate, see  Fig. \figref{fig:csrb_den_vor_a0}$(d_2)$. {However, some of the vortices and antivortices form vortex dipoles (vortex antivortex pair) or vortex pairs (pairs of identical charges) or vortex clusters, as depicted in Fig. \figref{fig:csrb_den_vor_a0}$(c_2)-(e_2)$. Without being annihilated, alongside the lone vortices and antivortices, these vortex dipoles, vortex pairs, and vortex or antivortex structures remain in the BEC for an extended period}.

When the paddle frequency $\omega_p$ increases, the vortex complexes exhibit a distinct behavior. At steady state, the number of antivortices vastly exceeds that of vortices for a CW rotation of the paddle potential with frequencies $\omega < \omega_p < 7\omega$. After removing the paddle potential, vortex-antivortex annihilation begins, finally eliminating all vortices from the condensate. In Fig. \figref{fig:vortex_number_a0}, we demonstrate  the number of vortices $(N^{A}_{+})$ and antivortices $(N^{\rm A}_{-})$ as a function of the rotation frequency $\omega_p$. The imbalance $\abs{N^{\rm A}_{+}-N^{\rm A}_{-}}$ is almost zero till $\omega_p = \omega$, indicating that an equal number of vortex-antivortex pairs are generated. Afterward, such imbalance gradually increases and becomes maximum at $\omega_p \approx 3\omega$ when only the antivortices (having negative circulation) exist in the system. For $\omega_p > 3\omega$, as it is evident from the Fig.\figref{fig:vortex_number_a0}, both total number of vortices, $N^{\rm A}_{+} + N^{\rm A}_{-}$ as well as the imbalance  decrease with $\omega_p$.

\begin{figure}[ht]
    \subfloat{\includegraphics[width=0.5\textwidth]{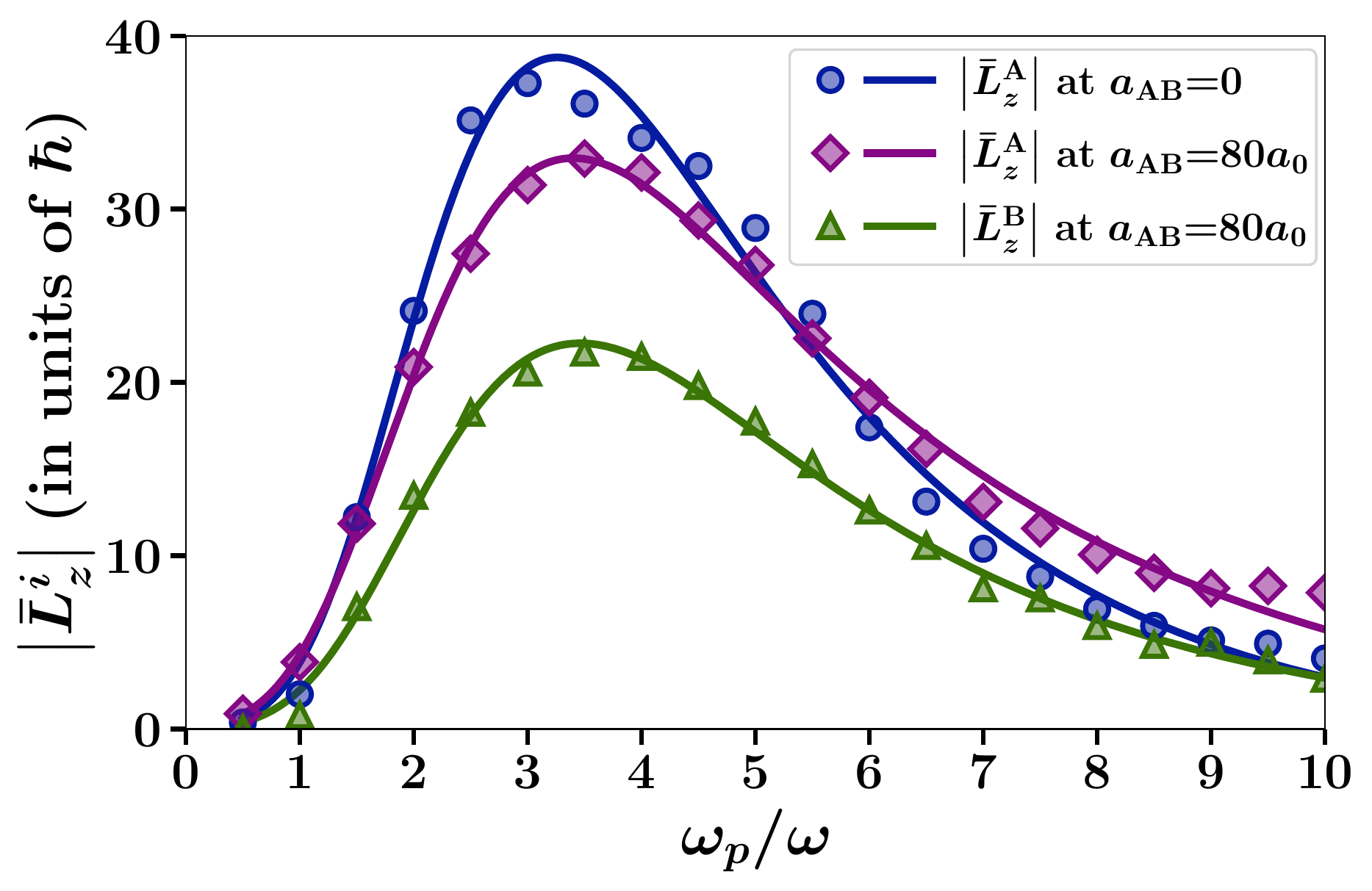}}
    \caption{Variation of the absolute value of time-averaged angular momentum $\abs{\bar{L}_z^{\rm A}}$ of species A at $a_{\rm AB}=0\text{ and }80a_0$, and $\abs{\bar{L}_z^{\rm B}}$ of species B at $a_{\rm AB}=80a_0$ as a function of paddle frequency (see the legends). The markers and solid curves show the values of angular momentum from the simulations and the fittings with skewed normal distribution, respectively. Here the paddle configuration is same as described in Fig. \figref{fig:csrb_den_vor_a0}.}
    \label{fig:csrb_ang_freq_a0_a80}
\end{figure}

Figures \figref{fig:csrb_vor_am_a0}$(a)$ and \figref{fig:csrb_vor_am_a0}$(b)$  show vorticity profiles, $\Omega_{\rm A}$, of species A  for  $\omega_p=3\omega$  and  $\omega_p=10\omega$, respectively,  at $t=3.5\rm s$.
Notably, the largest number of antivortices survives for $\omega_p = 3 \omega$  [Fig.\figref{fig:csrb_vor_am_a0}$(a)$] and this number falls as $\omega_p$ increases.
As $\omega_p$ increases beyond $\omega_p > 7\omega$, only a few of both vortices and antivortices survive due to a higher annihilation rate (per unit number of vortices-antivortices) of vortex-antivortex pairs [Fig. \figref{fig:csrb_vor_am_a0}(b)]. As a result, the system has almost no vortex or antivortex structure in the long-time dynamics (density profiles not shown here for brevity), imparting very less angular momentum to the condensate as shown in Figs. \figref{fig:csrb_vor_am_a0}(c) and \figref{fig:csrb_ang_freq_a0_a80}.

The above scenario of non-linear structure formations in species A can further be elucidated by invoking the angular momentum of species A, $L^{\rm A}_z$. The time evolutions of $L^{\rm A}_{z}(t)$ for various $\omega_p$ are displayed in Fig. \figref{fig:csrb_vor_am_a0}($c$). The $L^{\rm A}_z(t)$ remains negative throughout the time evolution, indicating the surplus of antivortices. For $\omega_p = \omega$, $L^{\rm A}_z(t)$ remains nearly constant within the ballpark. The $\abs{L^{\rm A}_z(t)}$ monotonically increases at the early stage of the dynamics and reaches a maximum at a time within the time interval $ \tau$, and then decreases to reach a stationary value in the long time dynamics. The maximum value of $\abs{L^{\rm A}_z(t)}$ is the largest for $\omega_p = 3\omega$, a result which emanates from the maximum number of antivortices displayed in Fig. \figref{fig:csrb_vor_am_a0}($a$). For larger $\omega_p$, the net angular momentum imparted to the condensate by the generated vortex-antivortex drastically diminishes, indicating a smaller imbalance between vortex and antivortex numbers [\figref{fig:csrb_vor_am_a0}($b$)].

\begin{figure}[ht]
    \subfloat{\includegraphics[width=0.5\textwidth]{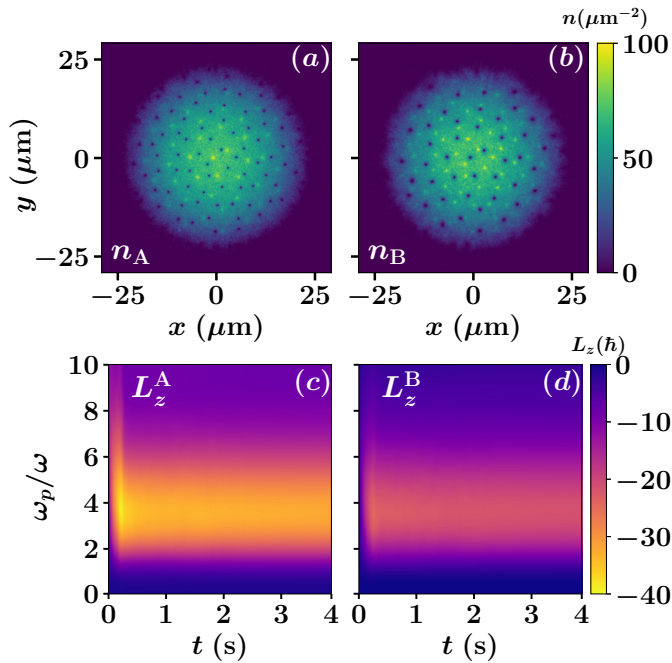}}
    \caption{Snapshots of density profiles ($(a),(b)$)  of species A $(n_{\rm A})$ and species B $(n_{\rm B})$, with interspecies scattering length $a_{\rm AB}=80a_0$ and $\omega_p=3\omega$ at $t=3.5\rm s$. Also shown are the variation of angular momentum $(c)$ $L_z^{\rm A}$ and $(d)$ $L_z^{\rm B}$ varying with paddle frequency $\omega_p$ and time $t$. The colorbars of \textit{top} and \textit{bottom} rows represent the number density $(n)$ in $\mu{\rm m}^{-2}$ and the angular momentum in units of $\hbar$ respectively.}
    \label{fig:csrb_den_ang_a80}
\end{figure}

The existence of paddle potential in species A affects species B for non-zero interspecies interactions $a_{\rm AB}$. For a strong enough interaction, the repulsive paddle potential on species A effectively acts as an attractive potential on species B. Due to rotation of this attractive potential vortex and antivortex are generated in species B \cite{jackson_2000_dissipation} (see Appendix \ref{sec:negative_paddle}). In particular, vortices and antivortices are created in species B, and their number can be controlled by $a_{\rm AB}$.
Additionally, the null density region at the vortex or antivortex site in one species is filled by the other species' localized density hump. Figure \figref{fig:csrb_den_ang_a80}($a$)-($b$) show the density pattern at $t=3.5\rm s$ for the interspecies interaction $a_{\rm AB}=80a_0$ and $\omega_p = 3 \omega$. Other parameters such as $\eta=0.05, d=0.1l \text{ and } V_0=10\mu_{\rm A}$ are similar to the $a_{\rm AB} =0$ case. Notably, the scattering lengths explored here ensure that the condensates are miscible, allowing us to directly analyze the role of mean-field coupling. Moreover, the paddle potential in species A performs CW rotation. Both species accommodate only antivortices solely in the long-time dynamics, which are similar to those of the non-interacting scenario. This behavior implies that within a particular frequency range, a cluster of identical vortices forms being entirely determined by the direction of paddle rotation, regardless of the species interaction. Furthermore, it is worth mentioning that species A possesses a smaller healing length due to the larger mass and the intraspecies interaction. This makes the vortices of species A smaller in size compared to those in species B.\par
The creation and stability of vortex complexes in the presence of non-zero interspecies interaction can be further comprehended by evaluating the angular momentum $L^{i}_z$ of both species. The time evolution of $L^{\rm A}_{z}$ and $L^{\rm B}_{z}$ as a function of $\omega_p$ are shown in Fig. \figref{fig:csrb_den_ang_a80}($c$) and Fig. \figref{fig:csrb_den_ang_a80}($d$), respectively.
A close inspection indicates that the angular momenta of both species are maximum at $\omega_p \approx 3.5 \omega$, similar to that in the $a_{\rm AB} = 0$ case [Figs. \figref{fig:csrb_den_ang_a80}(c)-(d), \figref{fig:csrb_ang_freq_a0_a80}]. For $\omega_p > 7\omega$, $L^{i}_z$ becomes very small due to the higher annihilation rate of the vortex-antivortex pairs. The $L^{\rm A}_z$ is more pronounced than  $L^{\rm B}_z$, indicating that the antivortex number is always high in species A. Most significantly, we find that the frequency response to the angular momentum follows skewed normal distribution
\bibnote[skew]{The skewed normal probability density function is given by,
$f(x)=\frac{2}{\sigma}\phi\qty(\frac{x-\mu}{\sigma})\Phi\qty(\alpha\frac{x-\mu}{\sigma})$
where $\phi(x)=e^{-x^2/2}/\sqrt{2\pi}$ is the standard normal probability density function and $\Phi(x)=\int_{\infty}^{x}\phi(t)\dd{t}$ is the cumulative distribution function. $\sigma,\mu$ and $\alpha$ are standard deviation, mean and skewed parameter, respectively. For $\alpha=0$, $f(x)$ becomes normal distribution.},
as depicted in the Fig.~\ref{fig:csrb_ang_freq_a0_a80}, with the maximum of the distribution occurring at $\omega_p \approx 3.2 \omega$ for the $a_{\rm AB} = 0$ and $\omega_p \approx 3.45 \omega$ for $a_{\rm AB} = 80a_{0}$.
Given that, at higher paddle frequencies ($\omega_p\gtrsim7\omega$), the annihilation process does not completely remove the vortices and antivortices from the condensates, leaving a small but finite angular momentum that leads to the long tail on the side of $\omega_p > 3\omega$, this distribution is quite expected.
Finally, let us comment that both species can end up with near-equal angular momenta the strong interaction limit $a_{\rm AB} \simeq 150a_{0}$. However, a detailed study of this regime is beyond the scope of the present manuscript.

The angular momentum achieves its maximum value for $\omega_p\approx3\omega\dash4\omega$ which can be explained by examining the sound velocity of the condensates. There are two distinct sound velocities for a binary BEC, namely, $c_+$ and $c_-$, representing the density and spin sound velocity, respectively \cite{eckardt_2004_groundstate,kim_2020_observation}. These two sound velocities can be expressed as
\smallskip 
\begin{align}
    c_{\pm}^2=\frac{1}{2}\qty[c_{\rm A}^2 + c_{\rm B}^2 \pm \sqrt{\qty(c_{\rm A}^2-c_{\rm B}^2)^2+4c_{\rm AB}^4}],
\end{align}
where $c_i=\sqrt{g_{ii}n_i/m_i}$ and $c_{\rm AB}^4=g_{\rm AB}^2n_{\rm A} n_{\rm B}/(m_{\rm A} m_{\rm B})$. $n_i$ represents peak density of the $i$-th condensate. For the non-interacting case with the peak density of $2.88\times10^{14}\rm{/cm}^3$, we have determined the sound velocity of species A (Cs) to be  $c_{\rm A}=2.36~\rm{mm/s}$ based on the averaged peak density $n_{\rm A}/2$, see also Refs. \cite{meppelink_2009_sound,kim_2020_observation}. The velocity of rotating paddle reads as $v = a\omega_p$ where $a=d/\eta$ is the semi-major axis of the paddle. With the values of parameters $d,\eta$ and $\omega_p=3\omega$ considered herein, we find that the paddle velocity amounts to $2.26~\rm{mm/s}$ which is very close to $c_{\rm A}$. This close proximity of paddle and sound velocities results in the maximum amount of angular momentum near the paddle frequency $\omega_p=3\omega$. For the interacting case with $a_{\rm AB}=80a_0$ we found the density sound velocity to be  $c_+=2.56\rm{mm/s}$ with the peak densities $n_{\rm A}=2.21\times10^{14}\rm{/cm}^3$ and $n_{\rm B}=2.27\times10^{14}\rm{/cm}^3$. The value of $c_+$ is very close the paddle velocity amounts to $2.63~\rm{mm/s}$ corresponding to the $\omega_p=3.5\omega$, where the absolute angular momenta of species A and B takes maximum values [Fig. \figref{fig:csrb_ang_freq_a0_a80}].
Note that vortex generation starts when the paddle rotation velocity $v$ surpasses a critical velocity which, in our case, is around $0.25c$, $c$ is the sound velocity. As already discussed, with the increase of paddle frequency, and hence $v$, a vortex-antivortex imbalance is created, increasing the angular momentum. When $v$ exceeds the sound velocity $c$ the drag-force becomes very pronounced resulting in stronger dissipation in the condensates \cite{frisch_1992_transition,carusotto_2006_bogoliubovcerenkov,jackson_2000_dissipation,ronning_2020_classical}. This dissipation causes the total vortex number and the vortex-antivortex imbalance to decrease, thus creating a peak of $\abs{L_z^{\rm A(\rm B)}} $ at $v\approx c$.

\subsection{Double Paddle}\label{sec:double_p}
After discussing the impact of a single paddle potential, we will look at a more complex scenario involving two paddle potentials. In this situation one can have two distinct scenarios depending upon the relative orientation of the paddle potentials. Here, we attempt to answer the question of how the addition of a second paddle potential and its relative rotational orientation relative to the first one alters the vortex structures and angular momentum of the system when compared to the case of a single paddle.
\begin{figure}[htbp]
    \includegraphics[width=0.48\textwidth]{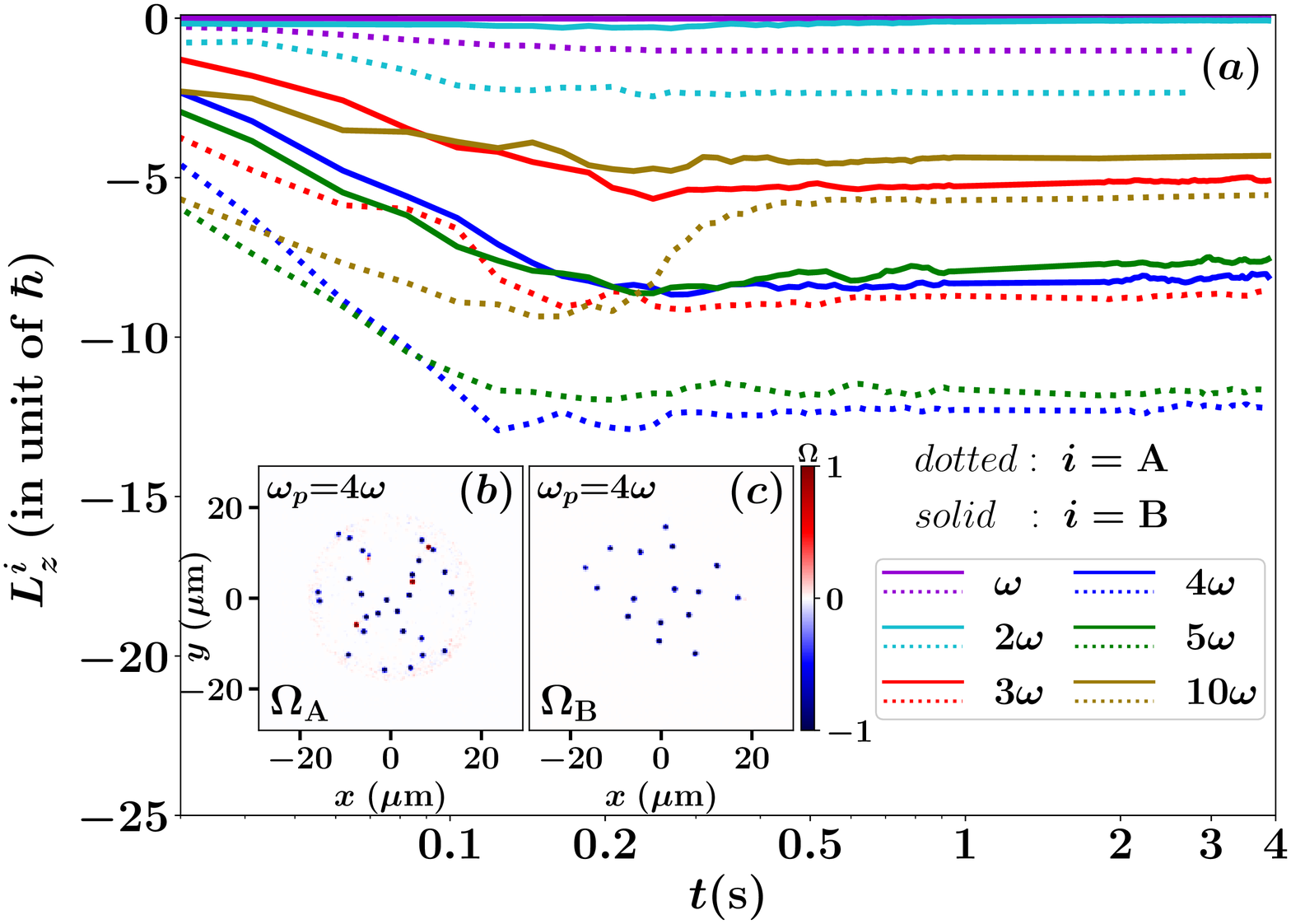}
    \caption{$(a)$ Time evolution ($\log$-scale) of angular momentum $L_z^{i}$ for the species $i=\rm A, B$ at interspecies interaction $a_{\rm AB}=80a_0$ for different paddle frequencies $\omega_p=\omega,2\omega,3\omega,4\omega$ and $10\omega$. Here two identical paddle rotates in species A in CW direction. The inset figures $(b)$ and $(c)$ depict the snapshot of vorticity profiles of species A ($\Omega_{\rm A}$) and species B ($\Omega_{\rm B}$), respectively, at $t=3.5\rm s$ with $\omega_p=4\omega$.}
    \label{fig:csrb_double_same_am_vor_a80}
\end{figure}
To that intent, we consider two paddles rotating in species A and having a center at $(\pm r_{\rm A}/4,0)$. Moreover, we choose $\eta=0.1$ and keep $d$ same as the single paddle case. Depending on the relative rotational orientation of the two potentials, different dynamical behavior can emerge. When both paddles rotate in the same direction, the effects are similar to those mentioned previously for a single paddle. To substantiate the above statement, we demonstrate the variation of angular momentum with time ($\log$-scale) in Fig. \figref{fig:csrb_double_same_am_vor_a80} for CW rotation of the paddle potential with interspecies scattering length $a_{\rm AB} = 80a_0$. For paddle frequency close to $\omega_{p} = 4 \omega$, $L^{\rm A}_z$ and $L^{\rm B}_z$ are most prominent, and the corresponding antivortex structures generated in species A and species B are shown in Fig. \figref{fig:csrb_double_same_am_vor_a80}$(b)$-($c$), respectively.  However, we should note that as we halved the paddle length with respect to the single paddle case and increased the paddle number to two, the maximum angular momentum generated in the system is reduced for the double co-rotating paddle than for the single paddle case. For example, $L^{\rm A}_z \approx -40 \hbar \omega$  at $\omega_p = 3\omega$ [Fig. \figref{fig:csrb_vor_am_a0}], whereas $L^{\rm A}_z \approx -9 \hbar \omega$ at the same $\omega_p$ [Fig. \figref{fig:csrb_double_same_am_vor_a80}] for the double paddle
    {\unskip\parfillskip 0pt \par}
\onecolumngrid

\begin{figure}[H]
    \includegraphics[width=\textwidth]{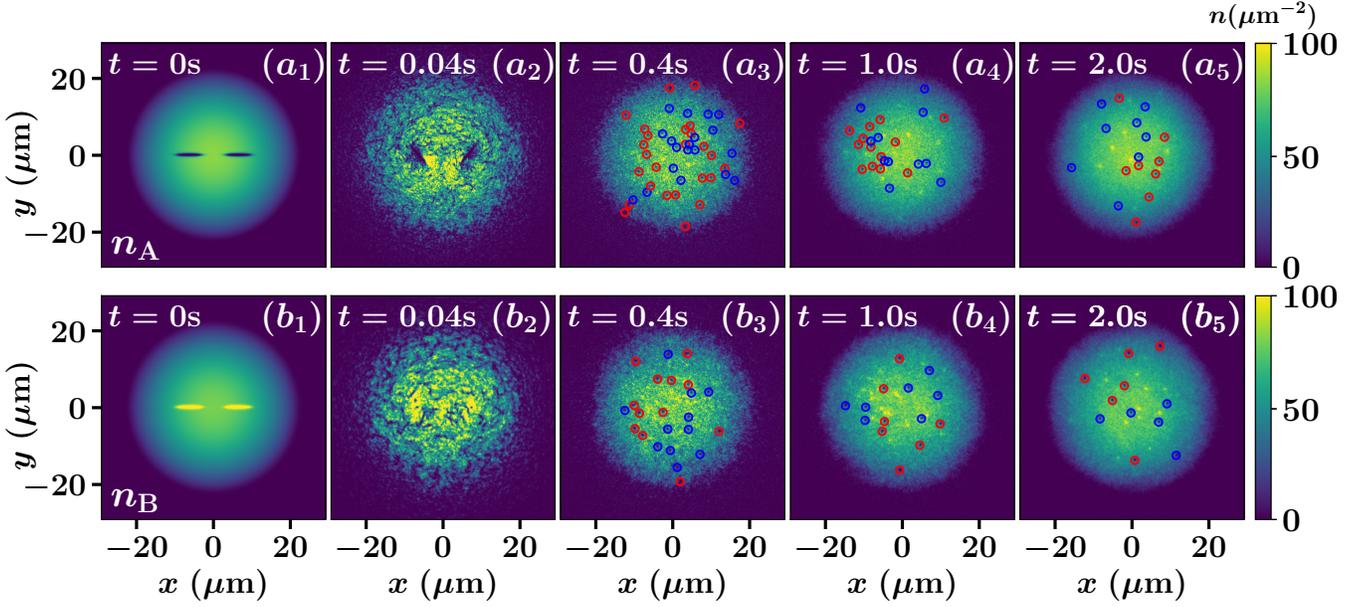}
    \caption{Snapshot of density of species A, $n_{\rm A}$ $((a_1)-(a_5))$ and species B, $n_{\rm B}$ $((b_1)-(b_5))$ with interspecies interaction $a_{\rm AB}=80a_0$ at different instants of time. Two elliptic paddles characterized by the parameters $\eta=0.1$ and $d=0.1l$ and rotating opposite to each other with frequency $\omega_p=5\omega$ within species A ($^{133}$Cs) are used to trigger the dynamics. The colorbars represent the number density $(n)$ in  $\mu\rm{m}^{-2}$. The vortices (antivortices) are marked with red (blue) in $(a_3)-(a_5)$ and $(b_3)-(b_5)$ (vortex identification is not done in $(a_2)$ and $(b_2)$ due to large number).}
    \label{fig:CsRb_a80_f5_double_opp}
\end{figure}
\twocolumngrid

\begin{figure}[H]
    \includegraphics[width=0.48\textwidth]{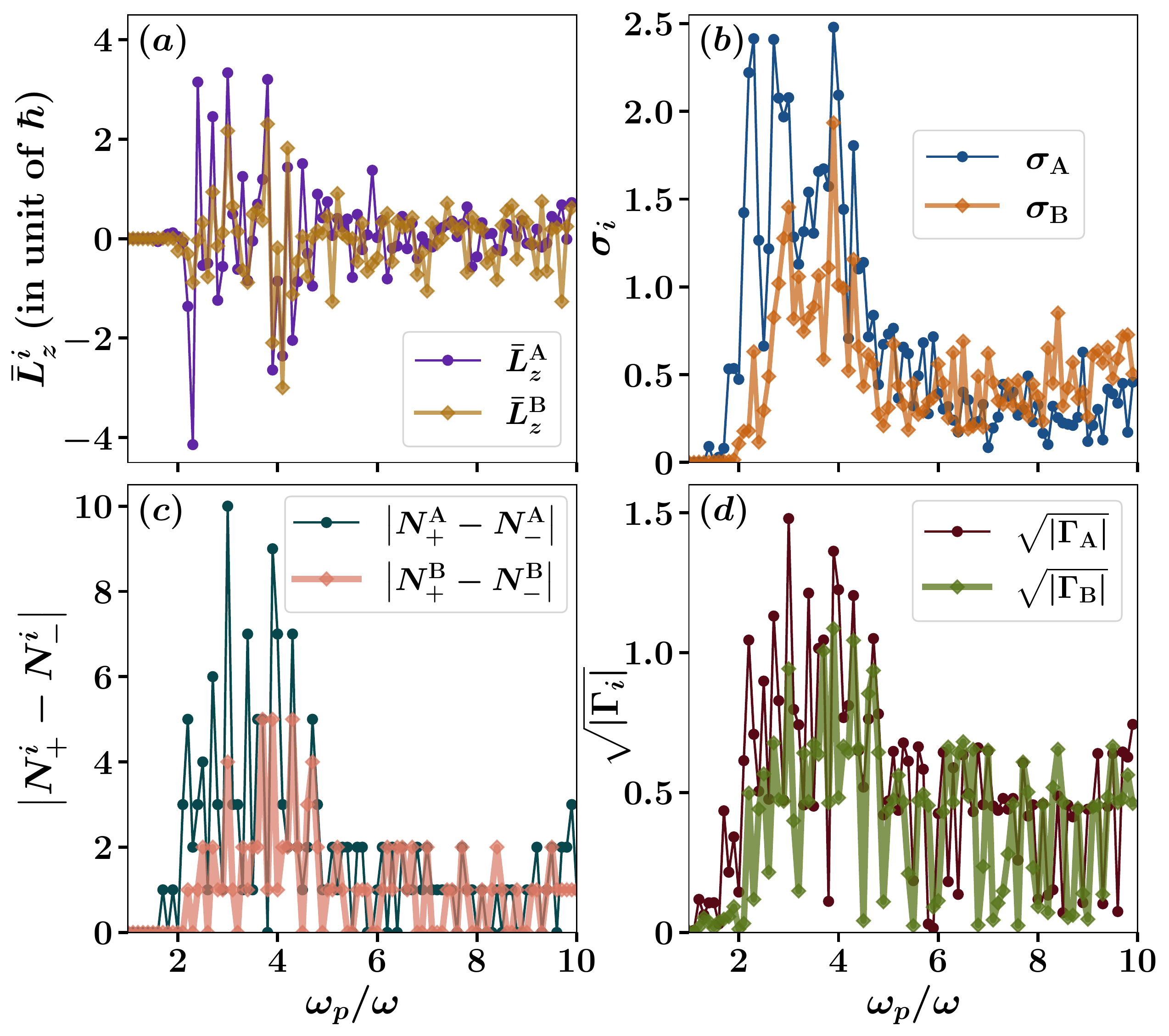}
    \caption{Variation of $(a)$ time-averaged angular momentum $\bar{L}_z^i$, its $(b)$ standard deviation $\sigma_i$ , $(c)$ vortex imbalance $\abs{N_+^i-N_-^i}$ at steady state, and $(d)$ $\sqrt{\abs{\Gamma_i}}$ ($\Gamma_i$ is the circulation at steady state),  as a function of $\omega_p$ for the $i$-th species ($i={\rm A,B}$). The scattering lengths are given by $a_{\rm AA} = 280a_{0}$, $a_{\rm BB} =100a_{0}$ and $a_{\rm AB} = 80a_{0}$. The dynamics has been triggered by employing two paddle potentials rotating counter-clockwise and the paddle configuration is described in Fig. \figref{fig:CsRb_a80_f5_double_opp}.}
    \label{fig:csrb_double_opp_am_std_dev}
\end{figure}
\noindent potentials. Surprisingly, a more interesting case occurs when one paddle rotates in the CW and the other in the CCW way [Fig.\figref{fig:CsRb_a80_f5_double_opp}($a_1$)-\figref{fig:CsRb_a80_f5_double_opp}($b_1$)].
Because the rotational directions of the paddles are opposite, each paddle contributes an equal number of vortices of the opposite sign, see Figs. \figref{fig:CsRb_a80_f5_double_opp}($a_3$) and \figref{fig:CsRb_a80_f5_double_opp}($b_3$). In the long-term dynamics of both species, this equal distribution of vortex and antivortex leads to a high rate of annihilation, meaning that just a few vortex, antivortex survive in the long-time dynamics [Figs.\figref{fig:CsRb_a80_f5_double_opp}($a_3$)-($a_5$) and \figref{fig:CsRb_a80_f5_double_opp}($b_3$)-($b_5$)].
To further appreciate the previous argument, we calculate the time average of the angular momentum defined as, $\bar{L}^{i}_z = \int L^{i}_z \dd{t}/\int \dd{t}$, for different rotation frequencies $\omega_p$ of the double paddle potentials, see Fig.~\ref{fig:csrb_double_opp_am_std_dev}(a). For $\omega_p < \omega$, the  $\bar{L}^{i}_z$ remains zero. Within $2 \omega \lesssim \omega_p \lesssim 5 \omega$, both $L_{\rm A}$ and $L_{\rm B}$  shows extremely fluctuating behaviour with respect to the $\omega_p$. Recall that this is also the frequency region where a maximum number of vortex-antivortex creations occur. The vortex-antivortex either annihilates each other or either of them drifts away from the condensate, leading to a finite imbalance of the vortex-antivortex number.
The finite imbalance between vortex antivortex numbers in the dynamics can result in the finite angular momentum of either positive or negative signs, a behavior which is highly fluctuating with respect to $\omega_p$. The fluctuation is somewhat reduced in the frequency range $\omega_p > 5 \omega$. Here, for increasing $\omega_p$, the annihilation mechanism becomes the dominant mechanism responsible for reducing both vortices and antivortices, and they exist in nearly equal numbers. Consequently, the imbalance between the vortex and antivortex number decreases, leading to a relatively small fluctuation in the $\bar{L}^{\rm A(B)}_{z}$.

Additionally,  we have calculated the standard deviation of time-averaged angular momentum $\bar{L}^{\rm A(B)}_z$ using data from five different runs with added noise for each run, see Fig.~\ref{fig:csrb_double_opp_am_std_dev}(b). The corresponding standard deviation $\sigma_{\rm A(B)}$ for species A (species B) is defined as,
\begin{align}
    \sigma^2_{\rm A(B)}=\frac{\sum_j{\qty(\bar{L}^{{\rm A(B)}, j}_{z} - \bar{L}^{\rm A(B)}_{z,\rm mean})^2}}{N_s}
\end{align}
where $\bar{L}^{\rm A(B)}_{z,\rm mean}=\sum_j{\bar{L}^{{\rm A(B)}, j}_{z}}/N_s$ and $N_s$ is number of data sets, each with different initial random noise. Fig.~\ref{fig:csrb_double_opp_am_std_dev}(b) depicts that the fluctuations are high in the frequency range $2 \omega < \omega_p < 5 \omega$. Furthermore, the fact that the fluctuations in $\bar{L}^{\rm A(B)}_z$ is indeed due to the fluctuations in the vortex-antivortex imbalance $\abs{N^{\rm A(B)}_{+} -N^{\rm A(B)}_{-}}$ can be evinced from the Fig.~\ref{fig:csrb_double_opp_am_std_dev}(c). Finally, before closing this section let us also remark on another interesting observation from our study that the $\abs{N^{\rm A(B)}_{+} -N^{\rm A(B)}_{-}}$ scales as $\sqrt{\abs{\Gamma_{\rm A(B)}}}$, where the quantity $\Gamma_{\rm A(B)} = \int {\Omega_{\rm A(B)}} \dd x \dd y$ represents the net circulation of the vortex clusters, see Fig.~\ref{fig:csrb_double_opp_am_std_dev}(d) where we demonstrate $\sqrt{\abs{\Gamma_{\rm A(B)}}}$ as a function of $\omega_p$.

\section{Energy spectra}\label{sec:energy_spec}

To better understand the system when it is subjected to a paddle potential, we compute its kinetic energy spectrum, whose scaling laws provide insights regarding the development of quantum turbulence in the system. Note that these scaling laws have already been well reported in the literature \cite{kobayashi_2005_kolmogorov,reeves_2012_classical,madeira_2019_quantum}. However, the primary objective here is to determine how the onset of turbulence depends on paddle frequencies or under what parameter regime the binary condensate system should develop turbulent features.

In order to do so we decompose the kinetic energy into compressible and incompressible parts associated with sound waves and vortices, respectively \cite{nore_1997_kolmogorov,white_2014_vortices}. The energy decomposition is performed by defining a density weighted velocity field, which reads $\sqrt{n_{i}}\vb{u}_{i}$ with $\vb{u}_{i} =\frac{\hbar}{m}\nabla \theta_i$, where  $n_i$ and $\theta_{i}$ are the position dependent condensate density and phase of the $i$-the species. The velocity field is separated into a solenoidal (incompressible) part $\vb{u}^{\rm ic}_i$ and a irrotational (compressible) part $\vb{u}^{\rm ic}_i$
such that $\vb{u}_{i} = \vb{u}^{\rm ic }_{i} + \vb{u}^{\rm c}_{i}$ and obeying $\div\vb{u}^{\rm ic}_i = 0$ and $\curl\vb{u}^{\rm c}_i = 0$. Once these velocity fields are calculated following the Refs. \cite{nore_1997_kolmogorov,horng_2009_twodimensional,mukherjee_2020_quench, ghoshdastidar_2022_pattern}, we can calculate incompressible energy $ (\mathcal{E}^{\rm ic}_i)$  and  compressible energy $ (\mathcal{E}^{\rm c}_i)$,
\begin{equation}
    \mathcal{E}^{\rm ic[c]}_i = \frac{1}{2} \int n_{i}\abs{\vb{u}^{\rm ic[c]}_{i}}^2 \dd{x} \dd{y}.
\end{equation}

Afterwards the compressible and incompressible energy spectra for the $i$-th species  can be calculated as
\begin{equation}
    E^{\rm ic[c]}_i(k) = \frac{k}{2} \sum_{q=x,y}^{}\int_{0}^{2 \pi}\abs{F_{q}(\sqrt{n_i}\vb{u}^{\rm ic [c]}_{q,i})}^2 \dd\phi,
\end{equation}
where $F_{q}(\sqrt{n_{i}}\vb{u}^{\rm ic[c]}_{i})$ denotes the Fourier transformation of $\sqrt{n_{i}\vb{u}^{\rm ic[c]}_{q,i}}$, corresponding to the $q$-th component of $\vb{u}_{i} = (u_{x,i}, u_{y, i})$.

\begin{figure}[t]
    \includegraphics[width=0.45\textwidth]{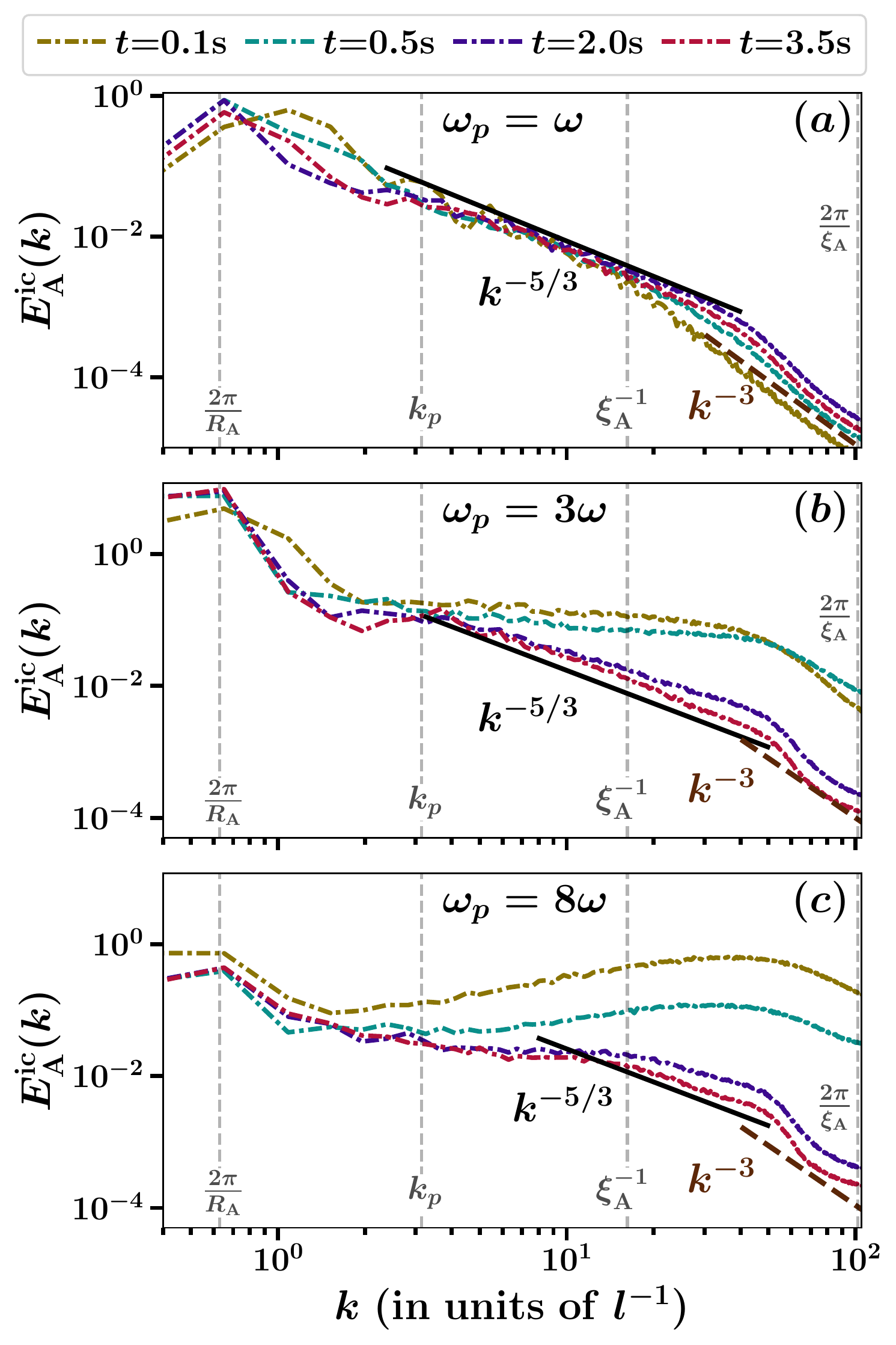}
    \caption{Incompressible kinetic energy spectra of species A, $E^{\rm ic}_{\rm A} (k)$, at different time instants (\textit{see legends}) for different paddle frequencies $\omega_p= (a)~\omega, (b)~3\omega$ and $(c)~8\omega$. The \textit{`solid'} and \textit{`dashed'} lines represent the slopes of $k^{-5/3}$ and $k^{-3}$, respectively. The interspecies scattering length $a_{\rm AB} = 0$. The dashed vertical lines define the positions at $k={2\pi}/{R_{\rm A}}$ ($R_{\rm A}=10l$ being the Thomas-Fermi radius), $k_p$ ($2\pi/a$, $a$ being the semi-major axis of the paddle), $\xi^{-1}_{\rm A}$ and $2\pi/\xi_{\rm A}$, respectively. Here the healing length $\xi_A$ of species A is $0.062l$. Except the $\omega_p$, all other parameters are the same as Fig.\figref{fig:csrb_den_vor_a0}.}
    \label{fig:kolmo_a0_Cs}
\end{figure}

We present incompressible energy spectra $E^{\rm ic}_{\rm A}(k)$ of species A in Fig. \figref{fig:kolmo_a0_Cs} at various time instants and frequencies $\omega_p$ corresponding to the single paddle case at $a_{\rm AB}=0$. Due to no interspecies interaction, species B is not impacted by the paddle potential, which allows us to focus on species A. For $\omega_{p} = \omega$, $E^{\rm ic}_{\rm A}(k)$ attains a stationary state at early time ($t=0.1 {\rm s}$) and maintains it till $t = 3.5{\rm s}$, as evidenced from the Fig. \figref{fig:kolmo_a0_Cs}($a$). Moreover, $E^{\rm ic}_{\rm A}(k)$ exhibits $k^{-3}$ power-law in the region $30\lesssim k \lesssim 100$ and $k^{-5/3}$ power law in the region $2\lesssim k \lesssim 30$. The $k^{-5/3}$ and $k^{-3}$ power laws are associated with the inertial range of energy cascade and internal structure of vortex core, respectively \cite{bradley_2012_energy,mithun_2021_decay}. Note that for $\omega_p = \omega$ vortex pairs and vortex dipole are noticed in Fig. \figref{fig:csrb_den_vor_a0}$(c_2)$-$(e_2)$ \cite{gauthier_2019_giant,reeves_2013_inverse,simula_2014_emergence,groszek_2016_onsager}. Surprisingly for the frequency $\omega_p =  3\omega$, where only the same sign vortices dominate, we notice that $k^{-3}$ spectrum develops for a very narrow $k$-range in our system, see Fig. \figref{fig:kolmo_a0_Cs}($b$) , and after that ($\omega_p>3\omega$) the spectra deviate from $-3$ scaling law, see Fig. \figref{fig:kolmo_a0_Cs}($c$) . However, while $k^{-5/3}$  spectrum develops in long time dynamics for a wide $k$-range, it does not emerge in early time dynamics [Fig. \figref{fig:kolmo_a0_Cs}($b$)]. Finally, we notice that  the $k$-ranges where the spectra follow $-5/3$ scaling become narrower with increase of $\omega_p$, see Fig. \figref{fig:kolmo_a0_Cs}($c$). This behaviour is expected since the system at $\omega_p \gtrsim 7 \omega$  is primarily governed by the generation of a huge number of sound waves caused by the rapid annihilation of the vortices and antivortices.
Another interesting observation from our study is that the most extended inertial range of the energy cascade occurs at the paddle frequency $\omega_p\approx3\omega$ where both species hold the maximum amount of angular momentum. The positions of the inertial ranges vary depending on $\omega_p$. For low $\omega_p (\simeq \omega)$ and high $\omega_p (> 5\omega)$ the inertial ranges occur respectively at lower and higher wavenumbers than the inverse of the healing length $(\xi_{\rm A}^{-1})$. For the intermediate frequencies it occurs at both sides of $k=\xi_{\rm A}^{-1}$, see Fig. \figref{fig:kolmo_a0_Cs}.

\begin{figure}[htbp]
    \includegraphics[width=0.45\textwidth]{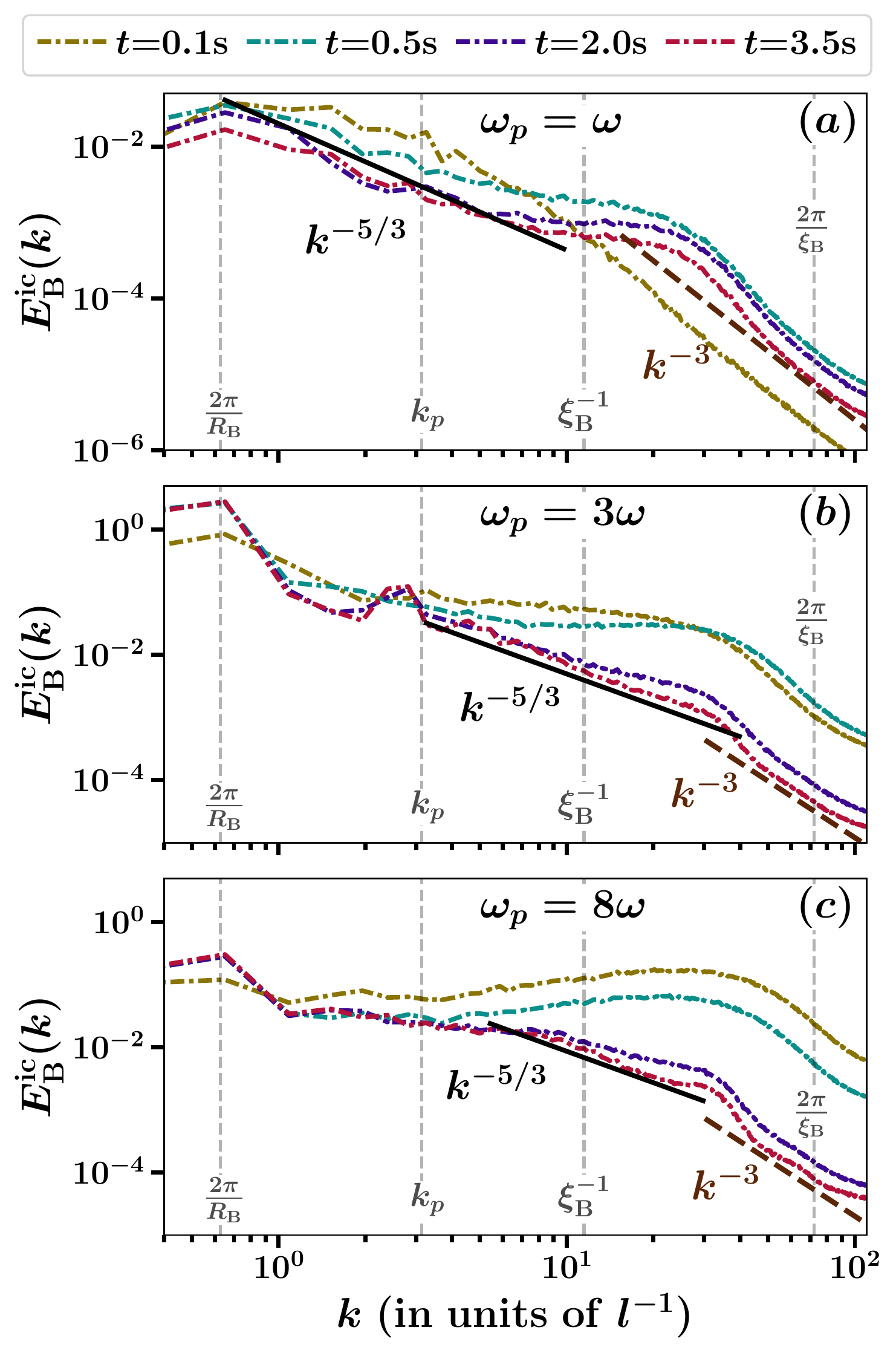}
    \caption{Incompressible kinetic energy spectra of species B, $E^{\rm ic}_{\rm B}(k)$,  at different time instants (\textit{see legends}) for different paddle frequencies $\omega_p= (a)~\omega, (b)~3\omega$ and $(c)~8\omega$. The \textit{`solid'} and \textit{`dashed'} lines represent the slopes of $k^{-5/3}$ and $k^{-3}$, respectively. The interspecies scattering length $a_{\rm AB} = 80a_0$. The dashed vertical lines define the positions at $k={2\pi}/{R_{\rm B}}$ ($R_{\rm B}\approx10l$ being the Thomas-Fermi radius), $k_p$ ($2\pi/a$, $a$ being the semi-major axis of the paddle), $\xi^{-1}_{\rm B}$ and $2\pi/\xi_{\rm B}$. Here the healing length $\xi_A$ of species B is $0.099l$. Except the $\omega_p$, all other parameters are same as Fig.\figref{fig:csrb_den_ang_a80}. }
    \label{fig:kolmo_a80_Rb}
\end{figure}

Next, we turn to the scenario of finite interspecies interaction characterized by $a_{\rm AB} = 80a_{0}$ and investigate whether species B produces the power-law spectra in the incompressible sector of its energy, see Figs. \figref{fig:kolmo_a80_Rb}($a$)-($c$). We note that $k^{-5/3}$ and $k^{-3}$ power laws are manifested in a similar manner within the range $ 1\lesssim k \lesssim 10$ and $20\lesssim k \lesssim 100$, respectively, for $\omega_p=\omega$.
Like in species A, the ranges of the $-5/3$ scaling law in species B become narrower with the increase of paddle frequency after $\omega_p\approx3\omega$ and the positions of the inertial ranges changes with $\omega_p$.
Moreover, species B contains vortices with larger cores than that of species A. At large $\omega_p$ the high-momentum acoustic waves are less in species B compared to that in species A because of the reduced strength of paddle potential at lower interspecies interaction. This makes incompressible kinetic energy at high momentum more discernible in species B than species A. Consequently, $k^{-3}$ scaling law, which is related to the vortex core structure, appears in $E^{\rm ic}_{\rm B}(k)$ for the frequency range $\omega\lesssim\omega_p\lesssim 10\omega$.
We note that in this condition $(a_{\rm AB}=80a_0)$ the $E^{\rm ic}_{\rm A}(k)$ does not demonstrate different behaviour with regard to $\omega_p$ when compared to that of $a_{\rm AB} = 0$ case (hence not shown here for brevity).

\begin{figure}[htbp]
    \includegraphics[width=0.45\textwidth]{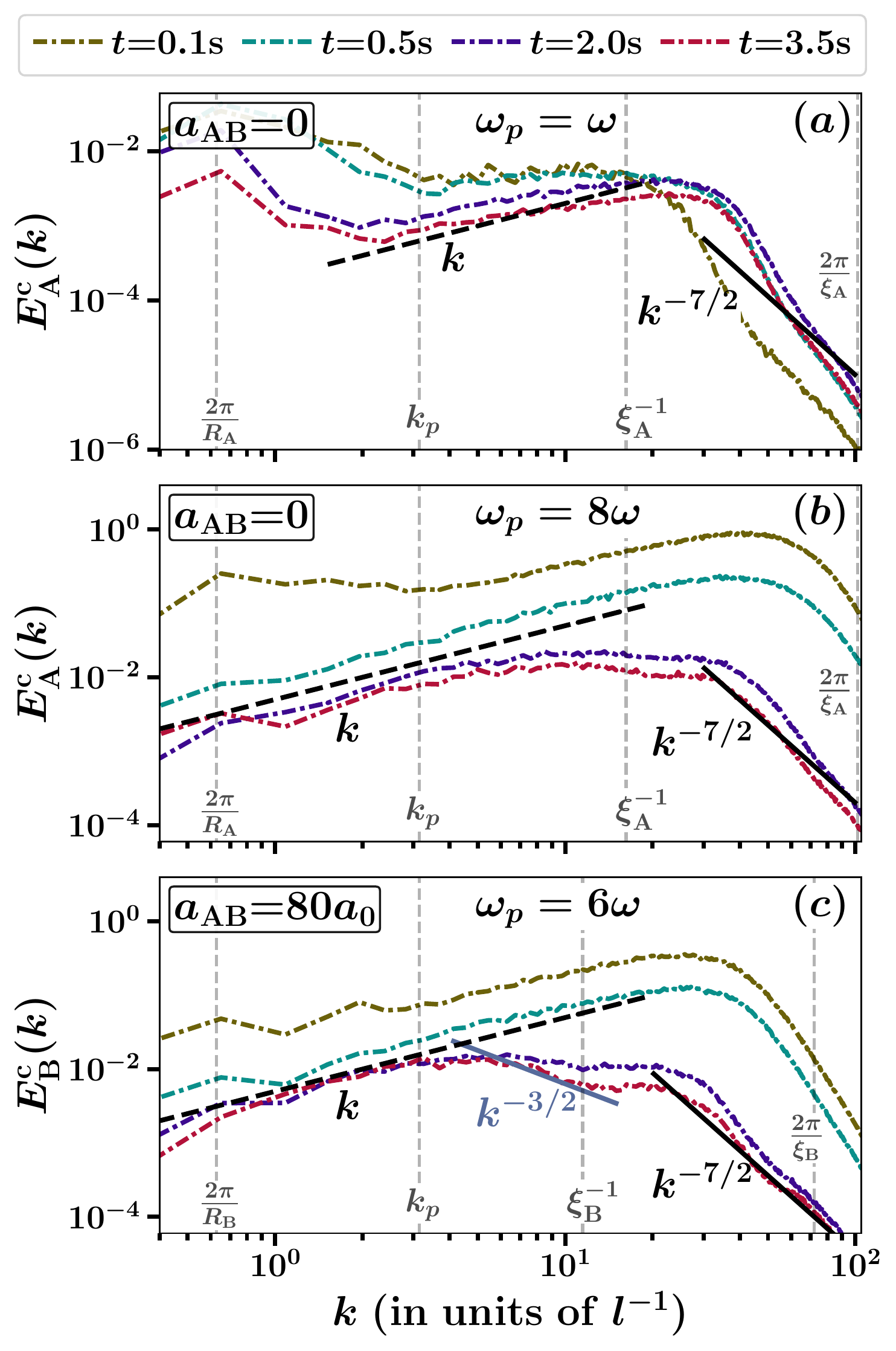}
    \caption{Compressible kinetic energy spectra of species A, $E^{\rm c}_{\rm A}(k)$ at $(a)$ $\omega_p=\omega$, $(b)~\omega_p=8\omega$ without interspecies interaction and of species B, $E^{\rm c}_{\rm B}(k)$ at $(c)~ \omega_p=6\omega$ with $a_{\rm AB}=80a_0$ at different time instants (\textit{see legends}). Black \textit{`solid'} and \textit{`dashed'} lines represent the slopes of $k^{-7/2}$ and $k$, respectively and blue \textit{`solid'} line in $(c)$ represents the slope of $k^{-3/2}$. The dashed vertical lines of $(a)-(b)$ and $(c)$ are described in Fig.\figref{fig:kolmo_a0_Cs} and \figref{fig:kolmo_a80_Rb} respectively. All the parameters are the same as discussed in Sec. \ref*{sec:single_p}. }
    \label{fig:kolmo_com}
\end{figure}

We now explain the compressible kinetic energy spectra \cite{mithun_2021_decay, ghoshdastidar_2022_pattern,shukla_2013_turbulence} $E^{\rm c}_{\rm A}(k)$ of species A and $E^{\rm c}_{\rm B}(k)$ of species B for a few representative cases subjected to the single paddle potential, shown in Fig. \figref{fig:kolmo_com}. To begin, in the case of $\omega_{p} = \omega$, we notice that a power-law region with $E^{\rm c}_{\rm A}(k) \propto k$ develops in the low-$k$ region of the spectrum, a relation that expresses the frequencies of Bogoliubov’s elementary excitations at low-wave number [Fig. \figref{fig:kolmo_a80_Rb}($a$)]. The spectrum reaches a maximum near $k$ ranges from 20 to 40 ( the peak positions differ for different time instants until the system reaches an equilibrium) before rapidly dropping. As the paddle frequency increases ($\omega_p\gtrsim7\omega$), the spectra $E^{\rm c}_{\rm A}(k)$ follows a power-law exponent of $-7/2$ at large $k$, as shown in Fig. \figref{fig:kolmo_com}(b) for a specific $\omega_p=8\omega$. Notably, this scaling is associated with superfluid turbulence of equilibrium sound waves, which has also been reported in Refs. \cite{mithun_2022_measurement,mithun_2021_decay,reeves_2012_classical}.
Interestingly enough, for $a_{\rm AB}=80a_0$, we observe the scaling law $k^{-3/2}$ in the intermediate $k$ range for the frequency $\omega_p \gtrsim 5 \omega$ as shown in Fig. \figref{fig:kolmo_com}(c) for $\omega_p=6\omega$. This power law which appears at $k$ higher than the driving wavenumber $k_p$ reveals the signatures of weak wave turbulence \cite{reeves_2012_classical,nazarenko_2006_wave}. Let us note that the acoustic disturbance must not be strong for the manifestation of this scaling \cite{nazarenko_2006_wave}; hence, it is more apparent in species B under weaker interspecies interaction regimes, while huge acoustic disturbances prevent the development of the same scaling in species A. We observed for strong enough interspecies interactions (\emph{e.g.} $a_{\rm AB}=140a_0$) the $-3/2$ scaling law disappears from species B (not shown). However, a detailed discussion of this is beyond the scope of the present manuscript.

\section{Conclusions}\label{conclusion}
We have explored the phenomenon of non-linear structure formations and their dynamics using optical paddle potential in a binary BEC composed of two distinct atomic elements. One of the species (species A) experiences rotating single or double paddle potentials, while the other species (species B) is only influenced via the interspecies contact interaction. The paddles are rotated for a finite amount of time, resulting in the creation of vortices. In long-time dynamics, the sign and number of the vortex are dependent on the frequency and orientation of paddle rotation.
Additionally, we discussed the effect of paddle rotation on other species. We observed many diagnostics to obtain insight into the dynamics, including density, vorticity, the $z$-component of the angular momentum, and the species' compressible and incompressible energy spectra.

Clusters of positive and negative vortices emerge within the system when a single paddle potential is rotated with a low rotational frequency.
Interestingly, when the frequency is gradually increased, we observe a transition to a regime dominated by same-sign vortices, with species A gaining the maximum angular momentum. At larger paddle frequencies, the annihilation of vortex-antivortex pairs becomes considerable, reducing the system's total vortical content.
The behavior mentioned above holds for species A both in the absence or presence of interspecies interaction. Interestingly enough, when interspecies contact is enabled, species B exhibits similar dynamical behavior.
However, species B has a substantially lower vortex and angular momentum than species A in the miscible regime. When two paddle potentials are employed, their relative orientation becomes crucial in determining the vortical content of species A.
For the paddles rotating with the same orientation, the behavior is almost identical to the single paddle applied to species A.
However, when the two paddles rotate opposite to each other, due to the almost equal number of vortex-antivortex structures formed regardless of the rotation frequency of the paddles, the net angular momentum imparted to the system during long-time dynamics fluctuates about zero.

Following that, we explored the system's dynamics by invoking the compressible and incompressible kinetic energy spectra. However, a key highlight of this work is its examination of various power-law scalings of the kinetic energy spectra.
We observed $-5/3$ and $-3$ power-law scaling in the low and high wavenumber regimes of the incompressible energy spectrum, respectively, in the low rotation frequency regime, where we saw clusters of identical sign vortices. These scalings provide evidence for the development of quantum turbulence in our system at low frequencies. However, analogous scaling is not apparent in the incompressible energy spectrum as the rotation frequency increases.

There are many research directions to be pursued as a future research endeavor. One straight would be to extend present work in the presence of finite temperature \cite{proukakis_2008_finitetemperature}. Extending the present work to the three-dimensional setup and exploring the corresponding non-linear defect formations would be equally interesting \cite{serafini_2017_vortex,cidrim_2017_vinen,xiao_2021_controlled,halder_2022_phase}. Another vital prospect would be to employ dipolar BEC to inspect the impact of the long-range interaction \cite{lahaye_2009_physics}. Finally, the investigations discussed previously would be equally fascinating at the beyond mean-field level, where significant correlations between particles exist \cite{cao_2017_unified}.

\section{Acknowledgment}
We thank the anonymous referees for their valuable comments that immensely improved the manuscript. We acknowledge National Supercomputing Mission, Government of India, for providing computing resources of ``PARAM Shakti" at Indian Institute of Technology Kharagpur, India.
\vspace*{-0.5cm}
\appendix
\renewcommand\thefigure{\thesection.\arabic{figure}}
\setcounter{figure}{0}
\section{Mass-Balanced binary Bose-Einstein condensate}\label{sec:equal_mass}

\begin{figure}[H]
    \includegraphics[width=0.48\textwidth]{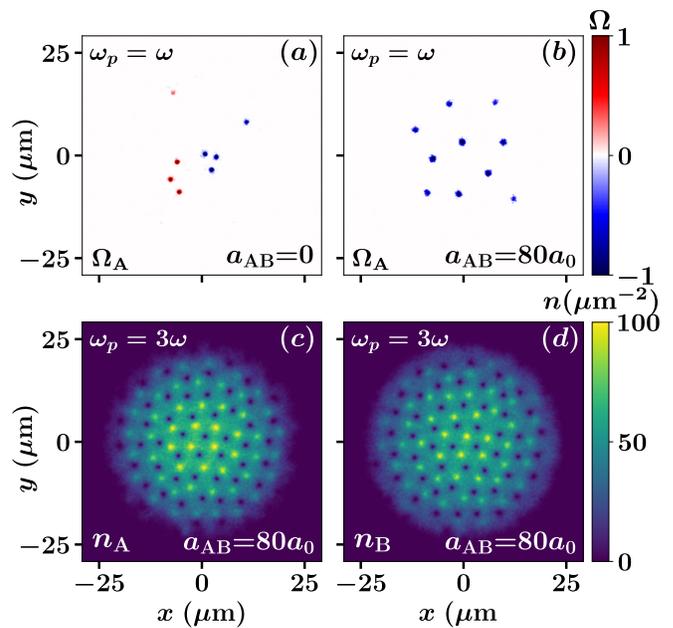}
    \caption{($a$)-($b$) Vorticity and ($c$)-($d$) density profiles of species A [$(a)$, $(c)$] and species B [$(b)$, $(d)$] at different paddle frequency, $\omega_p$, and different scattering lengths,  $a_{\rm AB}$, (see the legends) at $t=3.5\rm s$. The binary BEC is realized at two hyperfine levels of $^{87}$Rb atoms. To trigger the dynamics, an optical paddle potential is rotated in species A. The colorbars of \textit{top} and \textit{bottom} rows represent the vorticity ($\Omega$) and the number density ($n$) in $\mu\rm{m}^{-2}$.}
    \label{fig:RbRb_vor_den}
\end{figure}

In the main text, we have focused our discussion on the mass-imbalanced binary BECs, since such a system is the most suitable for creating species selective potential by the tune-out approach. To examine to what extent phenomenology differs from a system of mass-balanced system, here we consider a binary BEC composed of $^{87}$Rb atoms with two different hyperfine levels \cite{myatt_1997_production,mueller_2002_twocomponent}.
We take an equal number of atoms in both species, namely,  $N_{\rm A}=N_{\rm B}=60000$. The intra-species scattering lengths are $a_{\rm AA}=95.4a_0$ and $a_{\rm BB}=100.4a_0$ \cite{egorov_2013_measurement}. All other parameters, like trapping configuration and paddle configuration, are the same as the single paddle case of Sec. \ref{sec:vor_dyna}.
We examine the creation of vortices using a single rotating paddle with $a_{\rm AB}=0$ and $80a_0$. For paddle frequency $\omega_p=\omega$ we observe clustering of opposite sign vortices at $a_{\rm AB}=0$, see Fig. \figref{fig:RbRb_vor_den}(a). However, at higher interspecies scattering length $a_{\rm AB}=80a_0$, the clustering is not visible [Fig. \figref{fig:RbRb_vor_den}(b)], instead we observe a sparse cluster composed of same-signed vortices. The number of vortices organized into lattice structure increases as we increase the paddle frequency $(2\omega \lesssim \omega_p \lesssim 5\omega )$ in species A due to the direct impact of paddle rotation. And as an effect of interspecies interaction vortex lattice is also formed in species B [Fig. \figref{fig:RbRb_vor_den}(c)-(d)].

Since the interaction between two species in $^{87}$Rb-$^{87}$Rb are very similar, we find that the lattice-like structure that appeared here is more organised \cite{mueller_2002_twocomponent} than that of $^{133}$Cs-$^{87}$Rb binary BECs, compare Fig. \figref{fig:RbRb_vor_den}(c)-(d) with Fig. \figref{fig:csrb_den_ang_a80}(a)-(b).
Finally, we comment that we could not find any significant difference in incompressible and compressible kinetic energy spectra as a function of $\omega_p$ between the mass-balanced and mass-imbalanced systems [Fig. \figref{fig:kolmo_a0_Cs}(c)].

\section{Vortex Creation using negative paddle potential}\label{sec:negative_paddle}
\begin{figure}[H]
    \centering
    \includegraphics[width=\textwidth]{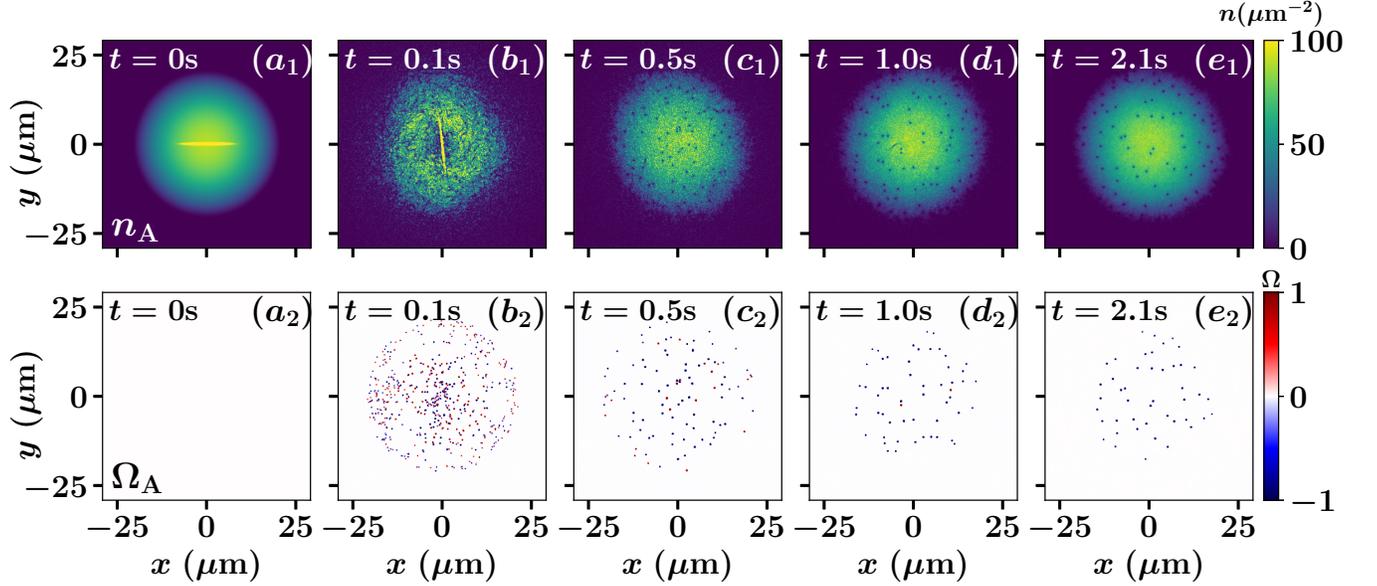}
    \caption{Snapshot of ($a_1$)-($e_1$) density $(n_{\rm A})$  species and ($a_2$)-($e_2$) vorticity $(\Omega_{\rm A})$ profiles of $^{133}$Cs at different instants of time (see legends). An elliptical paddle potential with amplitude $V_0=-10\mu_{\rm A}$ and characterized by the parameters $\eta = 0.05$ and $d=0.1l$ is rotated with the angular frequency $\omega_{p} = 4\omega$ within the species A. The colorbars of \textit{top} and \textit{bottom} rows represent the number density ($n$) in  $\mu\rm{m}^{-2}$ and the vorticity ($\Omega$). The binary BECs are initialized in a two dimensional harmonic potential with frequency $\omega /(2 \pi) = 30.832$Hz, $\lambda=100$ and having following intra- and interspecies scattering lengths $a_{\rm AA}=280a_{0}$, $a_{\rm BB} = 100.4a_{0}$, and $a_{\rm AB} = 0$. The number of atoms for both the species are $N_{\rm A}=N_{\rm B}=60000$. }
    \label{fig:csrb_pot_negative}
\end{figure}

In the main text of the article, we have focused our discussion on the paddle potential with $V_0 = 10 \mu_{\rm A}$. This results in the density depleted region in the condensate and creates vortex-antivortex structures when set into rotation. We remark that a rotating negative paddle potential would result in similar dynamics generating vortex-antivortex structures during the dynamics. In order to demonstrate that we have considered a $^{133}$Cs-$^{87}$Rb condensate of $N_{\rm A} = N_{\rm B} = 60000$ particles confined in the harmonic trap with the frequency $\omega/(2 \pi) = 30.832 \rm Hz$ and the anisotropy parameter $\lambda = 100$, $a_{\rm AA} = 280 a_{0}$ and $a_{\rm BB} = 100.4a_{0}$ and $a_{\rm AB} = 0$ [Fig.~\ref{fig:csrb_pot_negative}]. The species A is subjected to the paddle potential with $V_0 = -10 \mu_{\rm A}$, creating the density hump at its center, whereas the species B is unaffected [Fig.~\ref{fig:csrb_pot_negative}($a_1$)]. To trigger the dynamics, the paddle is rotated at the frequency $\omega_p = 4\omega$. A huge number of vortices and antivortices can be noticed at $t = 0.1s$ [Fig.~\ref{fig:csrb_pot_negative}($b_1$)-($b_2$)]. Then the number of vortices significantly decreases as time progresses [Fig.~\ref{fig:csrb_pot_negative}($c_1$)-($c_2$), ($d_1$)-($d_2$)]. Finally, in the long time dynamics antivortices dominate in the system [Fig.~\ref{fig:csrb_pot_negative}($e_1$)-($e_2$)]. Kindly note that similar behavior has been observed for the positive paddle potential at $\omega_p =  3\omega$, as discussed in the main text. This suggests that the phenomenon takes place irrespective of the attractive or repulsive paddle potential.

\onecolumngrid


\twocolumngrid

\bibliographystyle{apsrev4-2}
\bibliography{turbulence.bib}

\begin{thebibliography}{111}
\bibitem [{ske()}]{skew}%
  \BibitemOpen
  \href@noop {} {}\bibinfo {note} {The skewed normal probability density
  function is given by, $f(x)=\frac {2}{\sigma }\phi \qty (\frac {x-\mu
  }{\sigma })\Phi \qty (\alpha \frac {x-\mu }{\sigma })$ where $\phi
  (x)=e^{-x^2/2}/\sqrt {2\pi }$ is the standard normal probability density
  function and $\Phi (x)=\int _{\infty }^{x}\phi (t)\dd {t}$ is the cumulative
  distribution function. $\sigma ,\mu $ and $\alpha $ are standard deviation,
  mean and skewed parameter, respectively. For $\alpha =0$, $f(x)$ becomes
  normal distribution.}\BibitemShut {Stop}%
\end{thebibliography}
\end{document}